\documentclass[twocolumn,amsmath,amssymb,pre]{revtex4}

\usepackage{graphicx}
\usepackage{color}
\usepackage{dcolumn}
\usepackage{amsmath}
\usepackage{CJK}
\usepackage{subfigure}
\usepackage{sidecap}

\usepackage{mathrsfs}
\usepackage{amsfonts}
\usepackage{indentfirst}
\usepackage{bm}

\newcommand{\be}{\begin{equation}}
\newcommand{\ee}{\end{equation}}
\newcommand{\bey}{\begin{eqnarray}}
\newcommand{\eey}{\end{eqnarray}}
\newcommand{\bw}{\begin{widetext}}
\newcommand{\ew}{\end{widetext}}

\newcommand{\ra}{\rangle}
\newcommand{\la}{\langle}

\newcommand{\ba}{\begin{array}}
\newcommand{\ea}{\end{array}}
\newcommand{\bi}{\begin{itemize}}
\newcommand{\ei}{\end{itemize}}
\newcommand{\bem}{\begin{enumerate}}
\newcommand{\eem}{\end{enumerate}}

\begin{document}

\title{Effects of the Dzyaloshinsky-Moriya interaction on nonequilibrium thermodynamics in
      the $XY$ chain in a transverse field}

\author{Qian Wang$^{1}$, Duo Cao$^{2}$,
        and H.~T.~Quan$^{3,4}$\footnote{ Electronic address: htquan@pku.edu.cn}}

\affiliation{$^{1}$Department of Physics, Zhejiang Normal University, Jinhua 321004, China  \\
 $^2$Department of Physics, Shanghai Normal University, Shanghai 200234, China \\
 $^3$School of Physics, Peking University, Beijing 100871, China\\
 $^4$Collaborative Innovation Center of Quantum Matter, Beijing 100871, China}

\date{\today}

\begin{abstract}
  We examine the effects of the Dzyaloshinsky-Moriya (DM) interaction on
  the nonequilibrium thermodynamics in an anisotropic $XY$ spin chain,
  which is driven out of equilibrium by a sudden quench of the control
  parameter of the Hamiltonian.
  By analytically evaluating the statistical properties of the work distribution
  and the irreversible entropy production, we investigate the influences of the DM interaction
  on the nonequilibrium thermodynamics of the system with different parameters at various temperatures.
  We find that depending on the anisotropy of the system and the temperature, the DM interaction may have
  different impacts on the nonequilibrium thermodynamics.
  Interestingly, the critical line induced by the DM interaction can be revealed via the properties of the
  nonequilibrium thermodynamics.
  In addition, our results suggest that the strength of the DM interaction can be
  detected experimentally by studying the nonequilibrium thermodynamics.
\end{abstract}

\maketitle

\section{Introduction}
 The recent discovery of the exact nonequilibrium fluctuation relations
 \cite{esposito,campisi,seifert,mukamel,tasaki,crooks,jarzynski1} has spurred
 a lot of activities in both theoretical \cite{cbfr,jarzynski,rkwc} and
 experimental \cite{sdsb,collin,toyabe,tbam} investigations of the nonequilibrium
 thermodynamics in classical and quantum systems.
 A remarkable feature of the fluctuation relations is that they hold true for
 arbitrarily far from equilibrium process.
 With the help of the fluctuation relations, important equilibrium information, such as the free energy
 difference, can be extracted from the nonequilibrium process.
 Moreover, the fluctuation relations provide deep insights to the understanding of the
 microscopic foundations of the Second Law of thermodynamics \cite{jarzynski,campisi,mcph}.
 On the other hand, motivated by a series of spectacular experiments in atomic physics \cite{mgom,tktw,choi},
 a considerable amount of efforts have been devoted, during the past two decades, to the study of
 the nonequilibrium dynamics in quantum many-body systems \cite{apks}.
 Therefore, it is quite natural to explore the nonequilibrium thermodynamics in quantum many-body systems.
 Actually, the nonequilibrium thermodynamics has been studied in various quantum many-body systems,
 such as the one-dimensional transverse Ising \cite{dorner,heyl,silva,silvac,fusco}
 and $XY$ models \cite{mzhong,paraan,tjgap}, the Dicke model \cite{silvab}, $XYZ$ and $XXZ$ spin
 chains \cite{emhb,luca}, harmonic chains \cite{silvac,acmm}, the impurity Kondo model \cite{yesp,abtj},
 the sine-Gordon model \cite{ssag,tpam}, optomechanical systems \cite{mbax}, and Luttinger liquids \cite{dora}.
 In particular, the influence of the quantum criticality on the properties of the work and entropy production
 during nonequilibrium-driven processes has been revealed.

 The interactions in spin models studied in the aforementioned works are restricted to
 the symmetric spin-spin interactions.
 However, it is known that the Dzyaloshinsky-Moriya (DM) interaction is
 often present in many low-dimensional magnetic materials
 (e.g., see Refs.~\cite{derzhko,bqliu} and references therein).
 The DM interaction was initially introduced by Dzyaloshinsky \cite{dzya} to
 explain the weak ferromagnetic behavior shown in the antiferromagnetic crystals.
 It was later proven by Moriya that the DM interaction arises from the spin-orbit coupling \cite{moriya}.
 It has already been known that a number of interesting and unconventional
 effects \cite{fgzh,oata,vedv,pesin} originate from the DM interaction.
 Moreover, the phenomena caused by the DM interaction have been
 observed in many experiments \cite{ased,zakeri}.

 Several studies investigate the role of the DM interaction using tools
 borrowed from the field of quantum information, such as fidelity and
 fidelity susceptibility \cite{bwang,wenlong}, entanglement
 entropy \cite{bwang,jafari1,jafari2,zhengda}, and quantum and classical correlations \cite{bqliu}.
 Moreover, the impacts of the DM interaction on the dynamical properties
 of the system have also been studied using the dynamic
 transverse spin structure factor \cite{derzhko} and the decoherence factor \cite{wwcheng}.
 The influences of the DM interaction on the dynamical phase transition \cite{azimi} and the performance of the
 quantum heat engines also have been investigated \cite{azimi1,azimi2}.
 Very recently, the effect of the DM interaction in an alternating field quantum $XY$ model has been explored
 in Ref.~\cite{roy}.
 However, the effects of the DM interaction on the nonequilibrium thermodynamics have not been explored so far.

 In this work, we theoretically investigate the influences of the DM interaction
 on the nonequilibrium thermodynamics in a one-dimensional anisotropic $XY$ spin chain.
 We analytically evaluate the statistical properties of the work
 distribution and the irreversible entropy production
 in the process of a sudden quench of the system's control parameter.
 We will examine the impacts of the DM interaction on the nonequilibrium thermodynamics, for different
 anisotropy parameters and temperatures.
 Moreover, since a new critical line is present when the strength of the DM interaction
 is above a threshold value \cite{derzhko,zhongM,siskens},
 it is, therefore, interesting to study whether the nonequilibrium thermodynamics can be used to detect this
 new critical line.

 The remainder of this article is organized as follows.
 In Sec.~\ref{Stwo}, we briefly review some key concepts of the nonequilibrium thermodynamics.
 The explicit expression of the work distribution and the irreversible entropy production are introduced.
 In Sec.~\ref{Sthree}, we discuss the model and its diagonalization.
 Moreover, the analytical results of the characteristic function of the work distribution are also presented.
 In Sec.~\ref{Sfour}, the effects of the DM interaction on the statistical properties of the work distribution
 and the irreversible entropy production are studied.
 Finally, discussion of our main results and a summary are provided in Sec.~\ref{Sfive}.

\section{Nonequilibrium thermodynamics} \label{Stwo}

 In this section we briefly introduce some key concepts in the nonequilibrium thermodynamics
 and discuss the formulism that will be used in the rest of this article.
 We consider a quantum system described by a Hamiltonian $H[\lambda(t)]$, where $\lambda(t)$ is
 a time-dependent externally controlled parameter, called the work parameter in the field of
 nonequilibrium thermodynamics \cite{jarzynski}.
 Initially, we assume that the system equilibrates with a heat bath at the inverse temperature $\beta$ for a fixed
 value of the work parameter$\lambda(t\leq 0)=\lambda_0$.
 Therefore, the initial state of the system is described by the Gibbs state
 \be \label{IGS}
    \rho_0(\lambda_0)=\frac{e^{-\beta H_0}}{\mathcal{Z}_0(\lambda_0)},
 \ee
 where $\mathcal{Z}(\lambda_0)=\mathrm{Tr}[e^{-\beta H_0}]$, is the partition function.
 At $t=0$ the system is detached from the heat bath, and work is done on the system as the work parameter
 is varied from its initial value $\lambda_0$ to a final value $\lambda(t=\tau)=\lambda_f$.
 The initial and final Hamiltonian can be decomposed as
 $H_0=\sum_n E_n(\lambda_0)|n(\lambda_0)\ra\la n(\lambda_0)|$ and  $H_f=\sum_m E_m(\lambda_f)|m(\lambda_f)\ra\la m(\lambda_f)|$,
 respectively, where $E_n(\lambda_0)$ [$E_m(\lambda_f)$] is the $n$th ($m$th) eigenenergy of the
 initial (final) Hamiltonian with the eigenstate $|n(\lambda_0)\ra$ [$|m(\lambda_f)\ra$].
 According to the definition of the quantum work \cite{tasaki}, the work distribution of this nonequilibrium
 process is given by \cite{talkner}
 \be \label{WPDF}
   P(W)=\sum_{n,m}p_n^0 p_{m|n}^\tau\delta\{W-[E_m(\lambda_f)-E_n(\lambda_0)]\},
 \ee
 where $p_n^0=e^{-\beta E_n(\lambda_0)}/\mathcal{Z}(\lambda_0)$ denotes the probability to find the system in the $n$th eigenstate of
 $H_0$ when making the initial energy measurement.
 While $p_{m|n}^\tau$ is the transition probability to find the system in the $m$th eigenstate
 of $H_f$ at time $\tau$ if it were in the $n$th eigenstate of $H_0$ at
 the initial moment of time, i.e., $p_{m|n}^\tau=|\la m(\lambda_f)|\hat{U}(\tau)|n(\lambda_0)\ra|^2$, where
 $\hat{U}(\tau)=\hat{\mathcal{T}}\exp\left\{-\frac{i}{\hbar}\int_0^\tau H[\lambda(t)]\right\}$
 with $\hat{\mathcal{T}}$ as the time ordering operator.

 Usually, it is more convenient to use the characteristic function of work, which is defined as the Fourier
 transform of the work distribution \cite{talkner}:
 \begin{align}
   \chi(u)&=\int P(W) e^{iuW}dW  \nonumber  \\
        &=\mathrm{Tr}\{[\hat{U}(\tau)e^{-iuH_0}]\rho_0(\lambda_0)
        [e^{-iuH_f}\hat{U}(\tau)]^\dag\}.  \label{CHF}
 \end{align}
 It is worth mentioning here that the characteristic function in Eq.~(\ref{CHF}) has a similar form to
 the Loschmidt echo (LE), as was pointed out in Ref.~\cite{silva}.
 As a measure of the degree of the irreversibility in quantum systems, the LE
 has been used widely in many fields (see, e.g., Refs.~\cite{prosen,goussev}
 and references therein).
 In particular, the link between the quantum criticality and the property of the LE has been
 studied in several works \cite{quan,rjhj,ubsb,ubad}.

 \begin{figure*}
  \centering
  \includegraphics[width=0.7\textwidth]{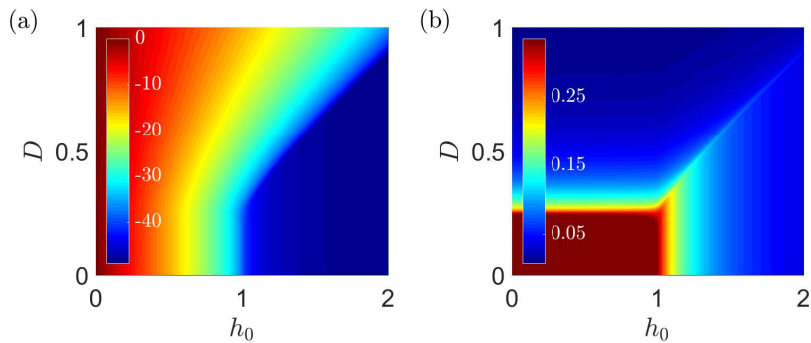}
  \caption{The averaged work (a) and the variance of the work distribution (b) as a function of $D$ and $h_0$.
   The parameters are chosen as follows: the perturbative quench of the magnetic field $\delta h=h_f-h_0=0.01$,
   the temperature $\beta=100$, the anisotropy $\gamma=0.5$, and the number of spins is $L=5000$.}
  \label{avgW}
 \end{figure*}

 The $n$th moment $\la W^n\ra$ and the cumulant $K_n$ of the work distribution
 can be obtained by expanding $\chi(u)$ and $\ln[\chi(u)]$, respectively.
 The final results are given by $\la W^n\ra=(-i)^n\left.\partial^n_u\chi(u)\right|_{u=0}$
 and $K_n=(-i)^n\left.\partial^n_u\ln[\chi(u)]\right|_{u=0}$ \cite{fusco}.
 Therefore, the averaged work $\la W\ra$ and the variance of the work distribution
 $\Sigma^2=\la W^2\ra-\la W\ra^2$ can be expressed as
 \be \label{AgVaW}
   \la W\ra=-i\partial_u\chi(u)|_{u=0},\
   \Sigma^2=(-i)^2\partial^2_u\ln[\chi(u)]|_{u=0}.
 \ee

 The statistical nature of the work in Eq.~(\ref{WPDF}) allows us
 to recast the form of the second law of thermodynamics as $\la W\ra\geq\Delta F$.
 This inequality implies the existence of an averaged irreversible work defined as
 $\la W_{\mathrm{irr}}\ra=\la W\ra-\Delta F$ \cite{jarzynski1,dorner,fusco,mzhong,paraan}.
 Then one can define the irreversible entropy production as
 \be \label{IRS}
   \Delta S_{\mathrm{irr}}=\beta\la W_{\mathrm{irr}}\ra=\beta(\la W\ra-\Delta F),
 \ee
 which quantifies the degree of the irreversibility of a nonequilibrium process.
 It has been proven that the irreversible entropy production can be expressed as the relative entropy
 between the instantaneous state $\rho_t$ at a given time of the evolution and the hypothetical
 Gibbs state $\rho_t^{eq}$ associated with the Hamiltonian of the system at that time
 \cite{dorner,fusco,utz,svaik},
 \be \label{RTS}
   \Delta S_{\mathrm{irr}}=S[\rho(t)||\rho_t^{eq}],
 \ee
 where $S(\rho||\kappa)=\mathrm{Tr}(\rho\log\rho-\rho\log\kappa)$ is the relative entropy between
 two arbitrary density matrices $\rho$ and $\kappa$.

 \begin{figure*}
  \centering
  \includegraphics[width=0.7\textwidth]{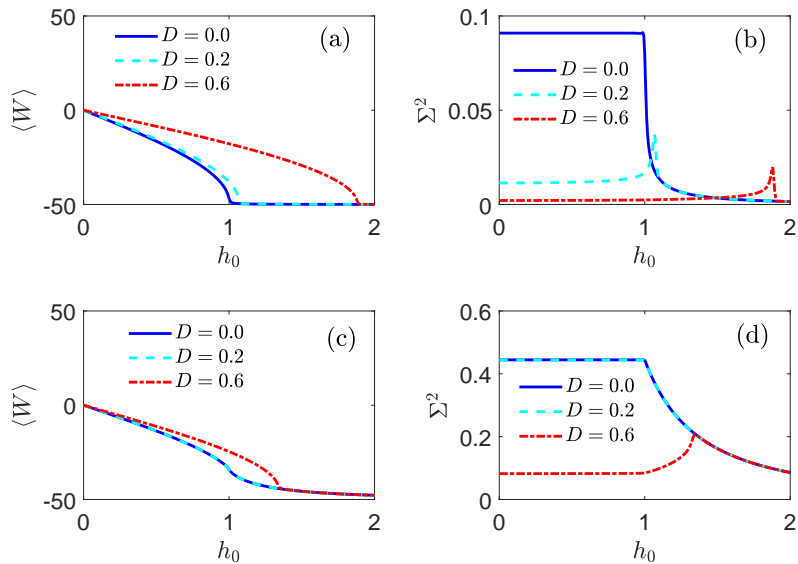}
  \caption{The averaged work as a function of $h_0$ under various $D$ for $\gamma=0.1$ (a) and $\gamma=0.8$ (c).
   The variance of the work distribution as a function of $h_0$ under various $D$ for
   $\gamma=0.1$ (b) and $\gamma=0.8$ (d).
   The perturbative quench of the magnetic field $\delta h=h_f-h_0=0.01$, the initial temperature
   and the number of spins are $\beta=100$ and $L=5000$, respectively.}
  \label{DpDg}
 \end{figure*}

\section{The $XY$ chain with the Dzyaloshinsky-Moriya interaction} \label{Sthree}

 The Hamiltonian of the one-dimensional anisotropic $XY$ chain in
 a transverse field with the Dzyaloshinsky-Moriya (DM)
 interaction reads \cite{derzhko,wwcheng,bqliu,roy}
 \begin{align} \label{XYDH}
  H_0=&-\sum_{j=1}^L\left(\frac{1+\gamma}{2}\sigma_j^x\sigma_{j+1}^x+\frac{1-\gamma}{2}\sigma_j^y\sigma_{j+1}^y
    +h_0\sigma_j^z\right) \nonumber \\
    &-\sum_{j=1}^L D\left(\sigma^x_j\sigma^y_{j+1}-\sigma^y_j\sigma^x_{j+1}\right),
 \end{align}
 where $L$ is the total number of spins, $\sigma_j^{x,y,z}$ are the Pauli matrices on the $j$th site,
 $\gamma (-1\leq\gamma\leq 1)$ is the anisotropy parameter, $h_0$ is the strength of the uniform
 external transverse magnetic field, and $D$ denotes the strength of the DM interaction along the $z$ direction.
 Here the periodic boundary condition $\sigma^{x,y,z}_{L+1}=\sigma^{x,y,z}_1$ is assumed.

 The Hamiltonian (\ref{XYDH}) can be diagonalized by mapping the spins to spinless fermions through
 the Jordan-Winger transformation \cite{lieb,sachdev}
 \be
  \begin{split}
   &\sigma_j^x=\prod_{l=1}^{j-1}(1-2f_l^\dag f_l)(f_j^\dag+f_j), \\
   &\sigma_j^y=-i\prod_{l=1}^{j-1}(1-2f_l^\dag f_l)(f_j^\dag-f_j), \\
   &\sigma_j^z=1-2f_j^\dag f_j.
  \end{split}
 \ee
 Here, $f_l (f_l^\dag)$ is the fermionic annihilation (creation) operator for the $l$th site.
 Then the Hamiltonian (\ref{XYDH}) can be rewritten as
 \begin{align}
   H_0=&-\sum_{j=1}^L\left[(1+2iD)f_j^\dag f_{j+1}+(1-2iD)f_{j+1}^\dag f_j\right. \notag \\
     &\left.+\gamma(f_j^\dag f_{j+1}^\dag+f_{j+1}f_j)+h_0(1-2f_j^\dag f_j)\right].
 \end{align}
 Now we transform the above Hamiltonian to the momentum space by employing the Fourier transformation
 $c_k=\frac{1}{\sqrt{L}}\sum_{j=1}^L f_je^{-ik j}, c_k^\dag=\frac{1}{\sqrt{L}}\sum_{j=1}^L f_j^\dag e^{ik j}$,
 where $c_k$ and $c_k^\dag$ are the fermionic annihilation and creation operators for the $k$th wave vector.
 The wave vectors $k$ are given by $k=2\pi m/L$ with $m=-(L-1)/2,\ldots,(L-1)/2$
 for an even $L$.
 Note that the lattice constant has been set as a unit.
 In terms of $c_k$ and $c_k^\dag$, the Hamiltonian can be rewritten as $H=\sum_k\mathcal{H}_k$,
 where $\mathcal{H}_k=\mathcal{C}_k^\dag\mathcal{M}_k\mathcal{C}_k$
 with $\mathcal{C}_k^\dag=(c_k^\dag, c_{-k})$ and
 \be
  \mathcal{M}_k=
  \begin{pmatrix}
    h_0-\cos k+2D\sin k  & -i\gamma\sin k  \\
    i\gamma\sin k  &  2D\sin k-h_0+\cos k
  \end{pmatrix}.\nonumber
 \ee

 \begin{figure}
  \includegraphics[width=\columnwidth]{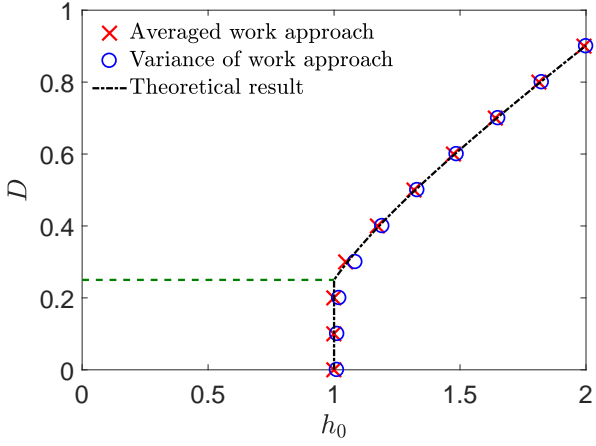}
  \caption{The location of the local minimums in the $D-h_0$ plane with $\gamma=0.5$, $\beta=100$,
   $\delta h=h_f-h_0=0.01$, and $L=5000$.
   The red crosses and the blue circles represent the local minimums in the derivatives of the averaged work
   and the variance of the work distribution with respect to the control parameter.
   The black dashed-dotted line denotes the critical point $h_c=1$ ($D<\gamma/2$) and the
   critical line $h_c=\sqrt{4D^2-\gamma^2+1}$ ($D>\gamma/2$). The dashed line indicates $D=\gamma/2$.}
  \label{newCP}
 \end{figure}

 The matrix $\mathcal{M}_k$ can be diagonalized by
 the following Bogoliubov transformation:
 \be \label{BGT}
   c_{\pm k}=\cos\left(\frac{\phi_k}{2}\right)\xi_{\pm k}
              \pm i\sin\left(\frac{\phi_k}{2}\right)\xi_{\mp k}^\dag,
 \ee
 where $\phi_k$ is given by
 \begin{align} \label{BAG}
  \phi_k=\arctan\left[\frac{\gamma\sin k}{h_0-\cos k}\right].
 \end{align}
 Here other fermionic operators $\{\xi_k,\xi_{-k}\}$ have been introduced through Eq.~(\ref{BGT}).
 The eigenvalues of $\mathcal{M}_k$ are
 $\lambda_\pm=\pm\varepsilon_k(h_0)+2D\sin k$ with $\varepsilon_k(h_0)=\sqrt{(h_0-\cos k)^2+\gamma^2\sin^2k}$.
 The final diagonalized form of the Hamiltonian (\ref{XYDH}) is given by
 $H_0=\sum_k\zeta_k(h_0)(2\xi_k^\dag\xi_k-1)$
 with $\zeta_k(h_0)=\varepsilon_k(h_0)+2D\sin k$ \cite{wwcheng,derzhko}.
 It is worth pointing out that only the energy spectrum $\zeta_k$ depends on $D$,
 while the coefficients of the Bogoliubov transformation and, hence, the energy eigenstates are independent of the
 strength of the DM interaction [cf.~Eq.~(\ref{BAG})].

 In our study, we consider the sudden quench process.
 Namely, the strength of the transverse magnetic field is changed suddenly at time $t=0$
 from the initial value $h_0$ to the final value $h_f$.
 Initially, the system is prepared in the thermal state with the inverse temperature $\beta$.
 Then the initial partition function in Eq.~(\ref{IGS}) reads
 \be
   \mathcal{Z}(h_0)=2^L\prod_k\cosh\left[\beta\zeta_k(h_0)\right].
 \ee
 Using an analogous diagonalization procedure outlined above, the post quench Hamiltonian
 can be rewritten in the diagonal form
 $H_f=\sum_k\zeta_k(h_f)(2\widetilde{\xi}_k^\dag\widetilde{\xi}_k-1)$,
 where $\{\widetilde{\xi}_k,\widetilde{\xi}_k^\dag\}$ are the post quench fermionic operators.
 Note that different values of the magnetic field imply different Bogoliubov transformations
 [cf.~Eqs.~(\ref{BAG}) and (\ref{BGT})].
 Therefore, the post quench fermionic operators $\{\widetilde{\xi}_k,\widetilde{\xi}_k^\dag\}$ are
 different from their pre quench counterparts $\{\xi_k,\xi_k^\dag\}$.
 However, these two classes of fermionic operators can be connected by inverting Eq.~(\ref{BGT}) for
 both the pre- and post quench fermionic operators. Their relations are \cite{wwcheng,dorner,mzhong,fusco,quan}
 \be \label{relation}
  \widetilde{\xi}_{\pm{k}}=\cos(\theta_k)\xi_{\pm{k}}\mp i\sin(\theta_k)\xi_{\mp{k}}^\dag,
 \ee
 where $\theta_k=(\widetilde{\phi}_k-\phi_k)/2$ with
 $\phi_k$ is given by Eq.~(\ref{BAG}) [$\widetilde{\phi}_k$ is given by Eq.~(\ref{BAG})
 with $h_0$ replaced by $h_f$].

 \begin{figure*}
  \centering
  \includegraphics[width=0.7\textwidth]{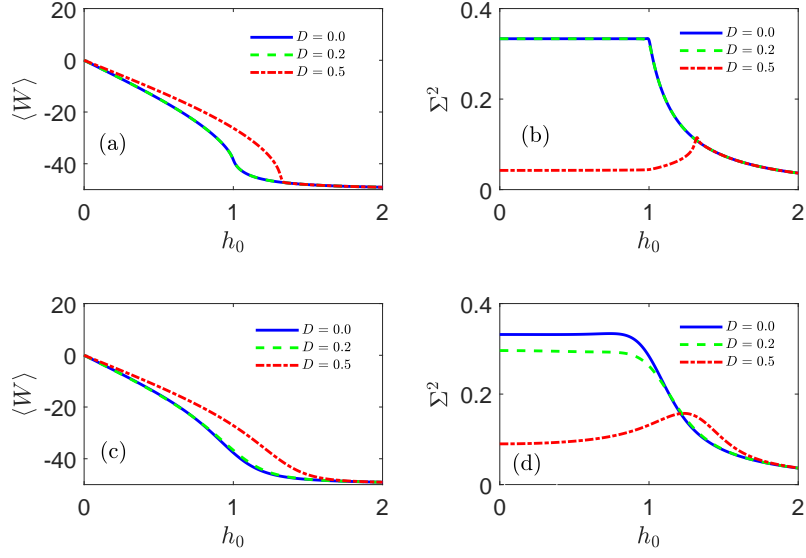}
  \caption{The averaged work as a function of $h_0$ for several
   different DM interactions at (a) $\beta=100$ and (c) $\beta=5$.
   The variance of the work distribution as a function of $h_0$ for different $D$ at (b) $\beta=100$
   and (d) $\beta=5$.
   The other parameters are $\gamma=0.5$, $L=5000$, and $\delta h=h_f-h_0=0.01$.}
  \label{Teffect}
 \end{figure*}

 In the sudden quench process, because $\hat{U}(\tau)$ equals the identity operator, the characteristic function
 in Eq.~(\ref{CHF}) can be simplified to
 \be \label{SPCH}
   \chi(u)=\frac{1}{\mathcal{Z}(h_0)}\mathrm{Tr}[e^{iuH_f}e^{-iuH_0}e^{-\beta H_0}].
 \ee
 The trace in Eq.~(\ref{SPCH}) can be evaluated analytically by employing the relations between the eigenstates
 of $H_0$ and $H_f$.
 To this aim, we first rewrite $H_0$ and $H_f$ in the following form
 \be \label{RWH}
   H_0=\sum_{k>0}H_k^0,\ H_f=\sum_{k>0}H_k^f,
 \ee
 where $H_k^0=[2\varepsilon_k(\hat{n}_k+\hat{n}_{-k}-1)+4D\sin k(\hat{n}_k-\hat{n}_{-k})],
 H_k^f=[2\widetilde{\varepsilon}_k(\hat{\widetilde{{n}}}_k+\hat{\widetilde{{n}}}_{-k}-1)
 +4D\sin k(\hat{\widetilde{{n}}}_k-\hat{\widetilde{{n}}}_{-k})]$
 with $\hat{n}_{\pm k}=\xi_{\pm k}^\dag\xi_{\pm k}
 (\hat{\widetilde{{n}}}_{\pm k}=\widetilde{\xi}_{\pm k}^\dag\widetilde{\xi}_{\pm k})$
 as the particle number operator of pre quench (post quench) fermions.
 The eigenstates of $H_k^0$ ($H_k^f$) are given by $|n_k,n_{-k}\ra$ ($|\widetilde{n}_k,\widetilde{n}_{-k}\ra$)
 with $n_{\pm k}=0,1$ ($\widetilde{n}_{\pm k}=0,1$).
 Then, from Eq.~(\ref{relation}), we express the eigenstates of $H_k^0$ in terms of the eigenstates of $H_k^f$
 as follows \cite{dorner,mzhong}:
 \be \label{RES}
  \begin{split}
  &|0_k,0_{-k}\rangle=\cos(\theta_k)|\widetilde{0}_k,\widetilde{0}_{-k}\rangle+
                     i\sin(\theta_k)|\widetilde{1}_k,\widetilde{1}_{-k}\rangle,  \\
  &|1_k,1_{-k}\rangle=\cos(\theta_k)|\widetilde{1}_k,\widetilde{1}_{-k}\rangle-
                     i\sin(\theta_k)|\widetilde{0}_k,\widetilde{0}_{-k}\rangle,   \\
  &|0_k,1_{-k}\rangle=|\widetilde{0}_k,\widetilde{1}_{-k}\rangle,\
   |1_k,0_{-k}\rangle=|\widetilde{1}_k,\widetilde{0}_{-k}\rangle.
  \end{split}
 \ee

 Combining Eqs.~(\ref{SPCH}), (\ref{RWH}), and (\ref{RES}),
 the characteristic function in Eq.~(\ref{SPCH}) then takes the form
 \begin{widetext}
 \begin{align}
  \chi(u)
  =&\frac{1}{\mathcal{Z}(h_0)}\prod_{k>0}\sum_{n=0,1} e^{-(iu+\beta)[2\varepsilon_k({n}_k
                  +{n}_{-k}-1)+4D\sin k({n}_k-{n}_{-k})]}
  \langle n_k,n_{-k}|e^{iu[2\widetilde{\varepsilon}_k(\widetilde{{n}}_k+\widetilde{{n}}_{-k}-1)
    +4D\sin k(\widetilde{{n}}_k-\widetilde{{n}}_{-k})]}|n_k,n_{-k}\rangle  \notag \\
  =&\frac{1}{\mathcal{Z}(h_0)}\prod_{k>0}\left\{e^{2(iu+\beta)\varepsilon_k}[e^{-2iu\widetilde{\varepsilon}_k}
  \cos^2(\theta_k)+e^{2iu\widetilde{\varepsilon}_k}\sin^2(\theta_k)]
  +e^{-2(iu+\beta)\varepsilon_k}[e^{-2iu\widetilde{\varepsilon}_k}
  \sin^2(\theta_k)+e^{2iu\widetilde{\varepsilon}_k}\cos^2(\theta_k)]\right. \notag \\
  &\left.+e^{4D\sin k(iu+\beta)}e^{-iu4D\sin k}+e^{-4D\sin k(iu+\beta)}e^{iu4D\sin k} \right\}. \label{AnCHF}
 \end{align}
 \end{widetext}
 With this analytical expression of the characteristic function of the work distribution,
 the Crooks relation and the Jarzynski equality can be easily verified \cite{mzhong,dorner}.
 Moreover, both the statistical moments of the work distribution and the irreversible entropy production
 can be exactly evaluated from the full expression of the characteristic function of the work distribution.

 \begin{figure}
  \includegraphics[width=\columnwidth]{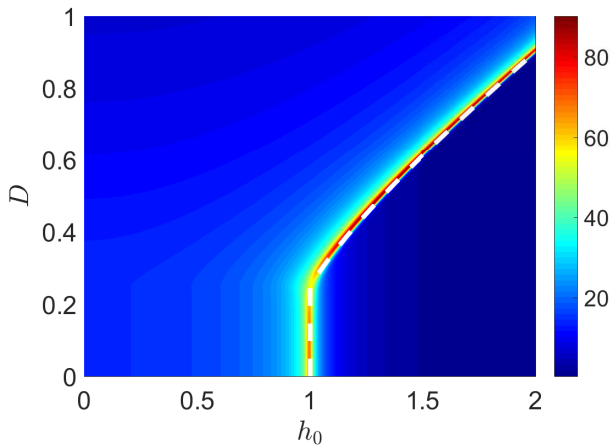}
  \caption{The irreversible entropy production as a function of $h_0$ and $D$ for
   perturbative quenches in the magnetic field with the amplitude
   $\delta h=h_f-h_0=0.01$ at a low temperature $\beta=100$.
   The other parameters are $\gamma=0.5$ and $L=5000$. The white dashed line indicates the
   critical point $h_c=1$ ($D<\gamma/2$), and the critical line $h_c=\sqrt{4D^2-\gamma^2+1}$
   ($D>\gamma/2$).}
  \label{IRepy}
 \end{figure}

\section{Work statistics and irreversible entropy production} \label{Sfour}

\subsection{Work statistics}
 Substituting the characteristic function in Eq.~(\ref{AnCHF}) into Eq.~(\ref{AgVaW})
 we get the averaged work and the variance of the work distribution
 \be \label{STAW}
   \la W\ra=\sum_{k>0}\la w\ra_k,\ \Sigma^2=\sum_{k>0}\sigma_k^2,
 \ee
 where
 \begin{align}
   &\la w\ra_k=\frac{2(h_0-h_f)\cos(\phi_k)\sinh(2\beta\varepsilon_k)}
           {\cosh(2\beta\varepsilon_k)+\cosh(4\beta D\sin k)},  \notag \\
   &\sigma_k^2=\frac{4(h_0-h_f)^2\cosh(2\beta\varepsilon_k)}{\cosh(2\beta\varepsilon_k)+\cosh(4\beta D\sin k)}
            -\la w\ra_k^2. \notag
 \end{align}
 Obviously, the expressions of $\la w\ra_k$ and $\sigma_k$ imply that the
 rescaled averaged work $\la W\ra/\delta h$ and the variance of the work distribution $\Sigma^2/\delta h^2$
 are independent of the amplitude of the quench strength.
 Moreover, the results in Ref.~\cite{paraan} for the $XY$ spin chain without the DM interaction can be
 recovered from Eq.~(\ref{STAW}) by setting $D=0$.
 Due to $\varepsilon_k$ and $\phi_k$ being independent of the value of $D$,
 the generalization to the case with the DM interaction involves only the modification of the
 denominator of $\la w\ra_k$ and $\sigma_k^2$, respectively.
 These analytical results allow us to evaluate the effects of the DM interaction on the nonequilibrium
 quantum thermodynamics in an exact way.

 \begin{figure*}
  \centering
  \includegraphics[width=0.7\textwidth]{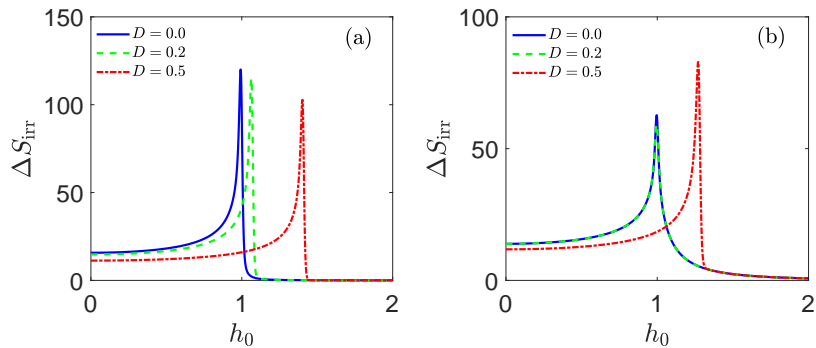}
  \caption{The irreversible entropy production as a function of $h_0$ at low temperature
   $\beta=100$ under various values of $D$
   with (a) $\gamma=0.1$ and (b) $\gamma=0.6$.
   The number of spins is $L=5000$, and the amplitude of the quench
   is $\delta h=h_f-h_0=0.01$.}
  \label{GamD}
 \end{figure*}

 In Figs.~\ref{avgW}(a) and \ref{avgW}(b), we plot the averaged work and the variance of the work
 distribution as a function of $h_0$ and $D$.
 Clearly, both $\la W\ra$ and $\Sigma^2$ are the continuous functions of $h_0$ and $D$.
 At low temperature, when the DM interaction is not strong enough, i.e., $D<\gamma/2$,
 the ground state is the ferromagnetic state, and the DM interaction does
 not have much influence on the ground sate.
 As a result, the behaviors of $\la W\ra$ and $\Sigma^2$ are insensitive to the DM interaction.
 However, due to the system changes from the gapped phase $(D<\gamma/2)$ to the
 gapless phase $(D>\gamma/2)$ at $D=\gamma/2$ \cite{zhongM,siskens}, the effects
 of the DM interaction emerge gradually when $D$ is larger than $\gamma/2$.

 In order to reveal more details about the influence of $D$ on the work distribution,
 in Fig.~\ref{DpDg} we plot the averaged work and the variance of the work distribution
 as a function of $h_0$ at $\beta=100$ under various $D$ and $\gamma$.
 The results in Fig.~\ref{DpDg} indicate that the DM interaction can increase the averaged work
 and suppress the fluctuation of the work distribution.
 From Fig.~\ref{DpDg}, we can also find that even though the behavior of the averaged
 work is similar for various $D$, the behavior of
 the variance of work distribution exhibits a remarkable change as $D$ increases.
 Specifically, when $D<\gamma/2$, the variance of the work distribution shows a sharp transition from a
 flat region to a monotonically decreasing region, and the DM interaction does not influence the variance of the work distribution.
 When $D>\gamma/2$, however, the variance of the work distribution shows a sharp peak.
 This phenomenon arises from the different properties between the
 gapped ($D<\gamma/2$) and the gapless ($D>\gamma/2$) phases,
 and confirms that the DM interaction substantially changes the
 variance of the work distribution.
 We note that Fig.~\ref{DpDg} shows also that the larger the values of $\gamma$, the weaker
 the influences of the DM interaction on the work distribution.

 It is well known that the discontinuities in the derivatives of the averaged work and/or the variance of
 work distribution with respect to the control parameter signal the presence of the quantum critical point
 \cite{fusco,paraan,silva,silvab,silvac,emhb,tjgap,luca}.
 On the other hand, it has already been verified that when $D<\gamma/2$ the critical point of system located at $h_c=1$,
 while a new critical line $h_c=\sqrt{4D^2-\gamma^2+1}$ appears when $D>\gamma/2$ \cite{zhongM,derzhko,siskens}.
 Therefore, an interesting question is whether the derivatives of the averaged work and/or the variance
 of the work distribution with respect to the control parameter can be used to reveal the critical line induced
 by the DM interaction.
 Notice that it has recently been found that the critical line mentioned above also
 exists in the open $XY$ model with DM interactions \cite{kos}.

 In Fig.~\ref{newCP}, we compare the numerical results which are the
 local minimums in the derivatives of the averaged
 work and the variance of the work distribution with the theoretical results.
 Obviously, a good agreement between the numerical and analytical results
 of the phase boundary can be seen in Fig.~\ref{newCP}.
 Our results, thus, imply that both the critical point and the critical line
 can be detected through the statistical properties of the work distribution.

 We further investigate the effects of the DM interaction on the work
 distribution under various temperatures.
 In Fig.~\ref{Teffect}, we show the averaged work and the variance of work distribution as a function
 of $h_0$ for several different DM interactions at $\beta=100$ and $\beta=5$, respectively.
 At low temperature, due to the aforementioned reason, the effect of the DM interaction on
 the statistical properties of the work distribution appear only when the value of $D$ satisfies $D>\gamma/2$.
 However, at high temperature, the sharp transitions in
 the averaged work and the variance of the work distribution are smoothed out due to the thermal fluctuations.
 Moreover, the high energy levels with large values of $k$ begin to play an important role in nonequilibrium dynamics of system.
 The DM interaction terms in the expressions of the averaged work
 and the variance of the work distribution [cf.~Eq.~(\ref{STAW})] become important.
 Thus, the evident influences of DM interaction on the statistical
 properties of the work distribution can be clearly seen even when $D<\gamma/2$ [cf.~Fig.~\ref{Teffect}(d)].
 The impacts of the DM interaction on the work distribution are, therefore,
 enhanced by increasing temperature (thermal fluctuations).

 \begin{figure*}
  \centering
  \includegraphics[width=0.7\textwidth]{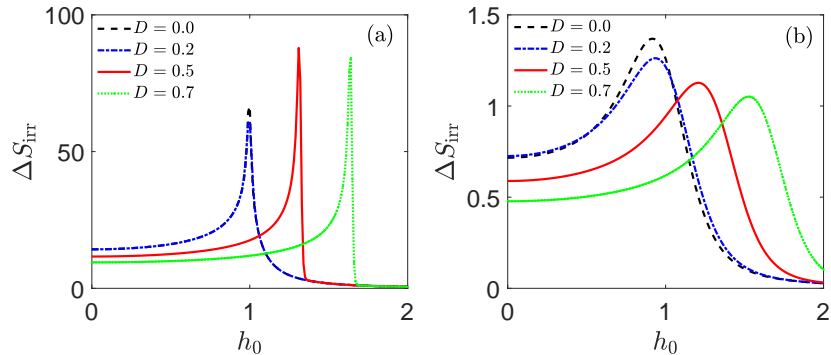}
  \caption{The irreversible entropy production as a function of $h_0$ for various $D$ at
   (a) $\beta=100$ and (b) $\beta=5$ with $\gamma=0.5$, $\delta h=h_f-h_0=0.01$,
   and the number of spins is $L=5000$.}
  \label{TemD}
 \end{figure*}

 \subsection{Irreversible entropy production}

 From the expression of the averaged work in Eq.~(\ref{STAW}), we find the irreversible entropy production
 in Eq.~(\ref{IRS}) is given by
 \be
   \Delta S_{\mathrm{irr}}=\beta\la W\ra+\sum_{k>0}\ln\left[\frac{\cosh(2\beta\widetilde{\varepsilon}_k)
                     +\cosh(4\beta D\sin k)}{\cosh(2\beta\varepsilon_k)+\cosh(4\beta D\sin k)}\right].
 \ee
 In Fig.~\ref{IRepy}, we plot the irreversible entropy production as a function of $D$ and $h_0$
 for $\delta h=h_f-h_0=0.01$ and $\gamma=0.5$ at low temperature $\beta=100$.
 Obviously, the irreversible entropy production is a continuous function of $D$ and $h_0$ and
 shows a sharp peak in the vicinity of the critical point or line.
 Therefore, the irreversible entropy production can be used to detect the critical line induced
 by the DM interaction.

 In the neighborhood of the critical regime, the sharp peak
 of $\Delta S_{\mathrm{irr}}$ can be explained by the quantum
 relative entropy between the instantaneous state at time $t$ and the hypothetical Gibbs state at that
 time [cf.~Eq.~(\ref{RTS})].
 For the case of a sudden quench, it has been verified that the relative entropy in Eq.~(\ref{RTS}) is equal to
 the distance between the initial and the final Gibbs states \cite{dorner,utz}.
 Therefore, in our study the irreversible entropy production can be rewritten as
 $\Delta S_{\mathrm{irr}}=S[\rho^{eq}(h_0)||\rho^{eq}(h_f)]$.
 Near the critical regime, a small change in the transverse field will lead
 to a dramatic change in the equilibrium state.
 Hence, the irreversible entropy production exhibits a sharp peak as the control parameter
 crosses the quantum critical point or line.
 For different values of the anisotropy parameters $\gamma$, the influences of the
 DM interaction on the behavior of the irreversible entropy production are shown in Fig.~\ref{GamD}.
%% \textbf{Such as in the work distribution case, the quantum phase transition occurring at $D=\gamma/2$ result in
%% the smaller the value of $\gamma$, the stronger the influence of the DM interaction
%% on the property of $\Delta S_{\mathrm{irr}}$.
%% Furthermore, we observe that the peak value of $\Delta S_{\mathrm{irr}}$
%% near the critical regime decreases with the increase of the DM interaction when the value of $\gamma$ is small.
%% However, for a large value of $\gamma$ the DM interaction can enhance the peak value of $\Delta S_{\mathrm{irr}}$.}

 We extend our study by investigating the effect of the DM interaction on the
 irreversible entropy production under various temperatures.
 In Fig.~\ref{TemD}, we plot the irreversible entropy production versus $h_0$ for various
 $D$ with $\beta$ equal to $100$ and $5$, respectively.
 First, we can see clearly that the thermal fluctuations generated
 by high temperature smooth out the sharp peak
 in the irreversible entropy production and suppress the value of
 $\Delta S_{\mathrm{irr}}$, regardless of the strength of the DM interaction.
 Second, as we mentioned above, the thermal fluctuations
 at high temperatures can enhance the effect of the
 DM interaction on the behavior of $\Delta S_{\mathrm{irr}}$.
 One can see from Fig.~\ref{TemD} that even when $D<\gamma/2$, the
 irreversible entropy production is different from that for $D=0$.

 As a last remark, we find that for a different size of the system the irreversible
 entropy production rescaled by the size of the system
 is an invariant quantity when the value of $D$ is fixed.
 Therefore, the irreversible entropy production scales extensively for a fixed D,
 i.e., $\Delta S_{\mathrm{irr}}\propto L$ as $L\to\infty$.
 On the other hand, since the large amplitude of the quench strength induces a
 large distance between the states of the pre- and post quench Hamiltonians,
 although $\Delta S_{irr}$ still has a peak near the critical point or line, its value
 is much larger than that for the small quench case.
 It is worth mentioning that at low temperature, for small quenches, the irreversible entropy
 production can be related to the fidelity susceptibility \cite{sharma,sptja}.

\section{Conclusions} \label{Sfive}

 In conclusion, we have studied the effects of the DM interaction on the nonequilibrium thermodynamics
 by exactly solving a quantum spin chain, i.e., the one-dimensional anisotropic $XY$ chain in a transverse field.
 The influence of the DM interaction on the nonequilibrium thermodynamics is revealed via the
 statistics of the work distribution and the irreversible entropy production.

 At low temperature, our results demonstrate that the nonequilibrium thermodynamics
 is insensitive to the DM interaction when the strength of the DM interaction
 is less than half of the anisotropy $\gamma$, i.e., $D<\gamma/2$.
 This originates from the fact that when $D<\gamma/2$ the ground state of the system is the
 ferromagnetic state, and the DM interaction does not have much influence on the ground state.
 However, due to the quantum phase transition at $D=\gamma/2$ \cite{zhongM,siskens}, the impacts of the DM interaction on
 the nonequilibrium thermodynamics will gradually emerge when $D>\gamma/2$.
 We find that the DM interaction can increase the averaged work while
 suppressing the fluctuation of the work distribution.
 Similarly, the DM interaction has nonvanishing effects on the irreversible entropy production
 only when $D>\gamma/2$.
 Interestingly, we find that the properties of the nonequilibrium thermodynamics
 can be used as a useful tool to detect the critical line induced by the DM interaction.
 At high temperature, due to the thermal fluctuations, the impacts of the DM interaction
 on the nonequilibrium thermodynamics can be enhanced by the thermal fluctuations.
 Therefore, a tiny change in the strength of the DM interaction will lead to a remarkable
 change in the properties of the nonequilibrium thermodynamics.

 Finally, we note that the procedure to experimentally determine the work statistics has been
 proposed \cite{dorner1,mazzola,mazzola1} and employed to measure the work distribution
 in a quantum system \cite{tbam}.
 In particular, the irreversible entropy production has been experimentally measured in Ref.~\cite{tbam1}.
 Therefore, our results suggest the possibility to experimentally probe the strength of the
 DM interaction in quantum systems by studying the nonequilibrium quantum thermodynamics.

 \acknowledgements

 H.T.Q. gratefully acknowledges support from the National Science Foundation of China
 under Grants No. 11775001 and No. 11534002, and the Recruitment
 Program of Global Youth Experts of China.


\begin{thebibliography}{99}

 \bibitem{esposito} E.~Esposito, U.~Harbol, and S.~Mukamel, Rev.~Mod.~Phys. {\bf 81}, 1665 (2009).

 \bibitem{campisi} M.~Campisi, P.~H\"{a}nggi, and P.~Talker, Rev.~Mod.~Phys. {\bf 83}, 771 (2011).

 \bibitem{seifert} U.~Seifert, Rep.~Prog.~Phys. {\bf 75}, 126001 (2012).

 \bibitem{mukamel} S.~Mukamel, Phys.~Rev.~Lett. {\bf 90}, 170604 (2003).

 \bibitem{tasaki} H.~Tasaki, arXiv:cond-mat/0009244v2.

 \bibitem{crooks} G.~E.~Crooks, Phys.~Rev.~E {\bf 60}, 2721 (1999).

 \bibitem{jarzynski1} C.~Jarzynski, Phys.~Rev.~Lett. {\bf 78}, 2690 (1997).

 \bibitem{cbfr} C.~Bustamante, J.~Liphardt, and F.~Ritort, Phys.~Today {\bf 58}(7), 43 (2005).

 \bibitem{jarzynski} C.~Jarzynski, Annu.~Rev.~Condens.~Matter Phys. {\bf 2}, 329 (2011).

 \bibitem{rkwc} R.~Klages, W.~Just, and C.~Jarzynski, editors,
               {\it Nonequilibrium Statistical Physics of Small Systems: Fluctuation
               Relations and Beyond}, (Wiley-VCH, Weinheim, 2013).

 \bibitem{sdsb} J.~Liphardt, S.~Dumont, S.~B.~Smith, I.~Tinoco~Jr., and C.~Bustamante, Science {\bf 296}, 1832 (2002).

 \bibitem{collin} D.~Collin, F.~Ritort, C.~Jarzynski, S.~B.~Smith, I.~Tinoco~Jr., and
                  C.~Bustamante, Nature (London) {\bf 437}, 231 (2005).

 \bibitem{toyabe} S.~Toyabe, T.~Sagawa, M.~Ueda, E.~Muneyuki, and M.~Sano, Nat.~Phys. {\bf 6}, 988 (2010).

 \bibitem{tbam} T.~B.~Batalhao, A.~M.~Souza, L.~Mazzola, R.~Auccaise, R.~S.~Sarthour, I.~S.~Oliveira, J.~Goold,
                G.~De~Chiara, M.~Paternostro, and R.~M.~Serra, Phys.~Rev.~Lett. {\bf 113}, 140601 (2014).

 \bibitem{mcph} M.~Campisi and P.~H\"{a}nggi, Entropy {\bf 13}, 2024 (2011).

 \bibitem{mgom} M.~Greiner, O.~Mandel, T.~Esslinger, T.~W.~Hansch, and I.~Bloch, Nature (Lodon) {\bf 415}, 39 (2002).

 \bibitem{tktw} T.~Kinoshita, T.~Wenger, and D.~S.~Wiss, Nature (London) {\bf 440}, 900 (2006).

 \bibitem{choi} J.-y.~Choi, S.~Hild. J.~Zeiher, P.~Schauss, A.~Rubio-Abadal, T.~Yefsah, V.~Khemani, D.~A.~Huse,
                I.~Bolch, and C.~Gross, Science {\bf 352}, 1547 (2016).

%% \bibitem{shen} P.~Jurcevic, H.~Shen, P.~Hauke, C.~Maier, T.~Brydges, C.~Hempel, B.~P.~Lanyon, M.~Heyl,
%%                R.~Blatt, and C.~F.~Roos, Phys.~Rev.~Lett. {\bf 119}, 080501 (2017).

 \bibitem{apks} A.~Polkovnikov, K.~Sengupta, A.~Silva, and M.~Vebgalattore, Rev.~Mod.~Phys. {\bf 83}, 863 (2011).

 \bibitem{heyl} M.~Heyl, A.~Polkovnikov, and S.~Kehrein, Phys.~Rev.~Lett. {\bf 110}, 135704 (2013).

 \bibitem{silva} A.~Silva, Phys.~Rev.~Lett. {\bf 101}, 120603 (2008).

 \bibitem{silvac} P.~Smacchia and A.~Silva, Phys.~Rev.~E {\bf 88}, 042109 (2013).

 \bibitem{fusco} L.~Fusco, S.~Pigeon, T.~J.~G.~Apollaro, A.~Xuereb, L.~Mazzola, M.~Campisi,
                 A.~Ferraro, M.~Paternostro, and G.~De Chiara, Phys.~Rev.~X {\bf 4}, 031029 (2014).

 \bibitem{dorner} R.~Dorner, J.~Goold, C.~Cormick, M.~Paternostro, and V.~Vedral, Phys.~Rev.~Lett. {\bf 109}, 160601 (2012).

 \bibitem{mzhong} M.~Zhong and P.~Q.~Tong, Phys.~Rev.~E {\bf 91}, 032137 (2015).

 \bibitem{paraan} F.~A.~Bayocboc. Jr. and F.~N.~C.~Paraan, Phys.~Rev.~E {\bf 92}, 032142 (2015).

 \bibitem{tjgap} T.~J.~G.~Apollaro, G.~Francica, M.~Paternostro, and M.~Campisi, Phys.~Scr. {\bf T165}, 014023 (2015).

 \bibitem{silvab} F.~N.~C.~Paraan and A.~Silva, Phys.~Rev.~E {\bf 80}, 061130 (2009).

 \bibitem{emhb} E.~Mascarenhas, H.~Braganca, R.~Dorner, M.~F.~Santos, V.~Vedral, K.~Modi, and J.~Goold,
                Phys.~Rev.~E {\bf 89}, 062103 (2014).

 \bibitem{luca} A.~De~Luca, Phys.~Rev.~B {\bf 90}, 081403(R) (2014).

 \bibitem{acmm} A.~Carlisle, L.~Mazzola, M.~Campisi, J.~Goold, F.~L.~Semiao, A.~Ferraro, F.~Plastina,
                V.~Vedral, G.~De~Chiara, and M.~Paternostro, arXiv:1403.0629v1.  %%%%%%%%%%%%%

 \bibitem{yesp} Y.~E.~Shchadilova, P.~Ribeiro, and M.~Haque, Phys.~Rev.~Lett. {\bf 112}, 070601 (2014).

 \bibitem{abtj} A.~Bayat, T.~J.~G.~Apollaro, S.~Paganelli, G.~De Chiara, H.~Johannesson, S.~Bose, and P.~Sodano,
                Phys.~Rev.~B {\bf 93}, 201016(R) (2016).

 \bibitem{ssag} S.~Sotiriadis, A.~Gambassi, and A.~Silva, Phys.~Rev.~E {\bf 87}, 052129 (2013).

 \bibitem{tpam} T.~Palmai, Phys.~Rev.~B {\bf 92}, 235433 (2015).

 \bibitem{mbax} M.~Brunelli, A.~Xuereb, A.~Ferraro, G.~De~Chiara, N.~Kiesel,
                and M.~Paternostro, New.~J.~Phys. {\bf 17}, 035016 (2015).

 \bibitem{dora} B.~Dora, A.~Bacsi, and G.~Zarand, Phys.~Rev.~B {\bf 86}, 161109(R) (2012). %%%%%%%%%%%%%**************

 \bibitem{derzhko} O.~Derzhko, T.~Verkholyak, T.~Krokhmalskii, and H.~B\"{u}ttner, Phys.~Rev.~B {\bf 73}, 214407 (2006).

 \bibitem{bqliu} B.~Q.~Liu, B.~Shao, J.~G.~Li, J.~Zou, and L.~A.~Wu, Phys.~Rev.~A {\bf 83}, 052112 (2011).

 \bibitem{dzya} I.~Dzyaloshinsky, J.~Phys.~Chem.~Solids {\bf 4}, 241 (1958).

 \bibitem{moriya} T.~Moriya, Phys.~Rev.~Lett. {\bf 4}, 228 (1960).

 \bibitem{fgzh} G.~F.~Zhang, Phys.~Rev.~A {\bf 75}, 034304 (2007).

 \bibitem{oata} O.~A.~Tretiakov and A.~Abanov, Phys.~Rev.~Lett. {\bf 105}, 157201 (2010).

 \bibitem{vedv} V.~E.~Dmitrienko and V.~A.~Chizhikov, Phys.~Rev.~Lett. {\bf 108}, 187203 (2012).

 \bibitem{pesin} D.~Pesin and L.~Balents, Nat.~Phys. {\bf 6}, 376 (2010).

 \bibitem{ased} I.~A.~Sergienko and E.~Dagotto, Phys.~Rev.~B {\bf 73}, 094434 (2006).

 \bibitem{zakeri} K.~Zakeri, Y.~Zhang, J.~Prokop, T.-H.~Chuang, N.~Sakr, W.~X.~Tang, and
                  J.~Kirschner, Phys.~Rev.~Lett. {\bf 104}, 137203 (2010).

 \bibitem{wenlong} W.~L.~You and Y.~L.~Dong, Phys.~Rev.~B {\bf 84}, 174426 (2011).

 \bibitem{bwang} B.~Wang, M.~Feng, and Z.~Q.~Chen, Phys.~Rev.~A {\bf 81}, 064301 (2010).

 \bibitem{jafari1} R.~Jafari, M.~Kargarian, A.~Langari, and M.~Siahatgar, Phys.~Rev.~B {\bf 78}, 214414 (2008).

 \bibitem{jafari2} M.~Kargarian, R.~Jafari, and A.~Langari, Phys.~Rev.~A {\bf 79}, 042319 (2009).

 \bibitem{zhengda} Z.~D.~Hu, Q.~L.~He, and J.~B.~Xu, Physica A {\bf 391}, 6226 (2012).

 \bibitem{wwcheng} W.~W.~Cheng and J.-M. Liu, Phys.~Rev.~A {\bf 79}, 052320 (2009).

 \bibitem{azimi} M.~Azimi, M.~Sekania, S.~K.~Mishra, L.~Chotorlishvili,
                 Z.~Toklikishvili, and J.~Berakdar, Phys.~Rev.~B {\bf 94}, 064423 (2016).

 \bibitem{azimi1} M.~Azimi, L.~Chotorlishvili, S.~K.~Mishra, T.~Vekua, W.~H\"{u}bner
                  and J.~Berakdar, New.~J.~Phys. {\bf 16}, 063018 (2014).

 \bibitem{azimi2} L.~Chotorlishvili, M.~Azimi, S.~Stagraczynski, Z.~Toklikishvili, M.~Sch\"{u}ler,
                  and J.~Berakdar, Phys.~Rev.~E {\bf 94}, 032116 (2016).

 \bibitem{roy} S.~Roy, T.~Chanda, T.~Das, D.~Sadhukhan, A.~Sen, and U.~Sen, arXiv:1710.11037v1.

 \bibitem{zhongM} M.~Zhong, H.~Xu, X.~X.~Liu, and P.~Q.~Tong, Chin.~Phys.~B {\bf 22}, 090313 (2013).

 \bibitem{siskens} T.~J.~Siskens, H.~W.~Capel, and K.~J.~F.~Gaemers, Physica A {\bf 79}, 259, (1975).

 \bibitem{talkner} P.~Talkner, E.~Lutz, and P.~H\"{a}nggi, Phys.~Rev.~E {\bf 75}, 050102(R) (2007).

 %%\bibitem{talkner1} P.~Talkner and P.~H\"{a}nggi, J.~Phys.~A {\bf 40}, F569 (2007).

 \bibitem{prosen} T.~Gorin, T.~Prosen, T.~H.~Seligman, and M.~Znidaric, Phys.~Rep. {\bf 435}, 33 (2006).

 \bibitem{goussev} A.~Goussev, R.~A.~Jalabert, H.~M.~Pastawski, and D.~A.~Wisniacki, Phil.~Trans.~R.~Soc.~A {\bf 374}:20150383 (2016).

 \bibitem{quan} H.~T.~Quan, Z.~Song, X.~F.~Liu, P.~Zanardi, and C.~P.~Sun, Phys.~Rev.~Lett. {\bf 96}, 140604 (2006).

 \bibitem{rjhj} R.~Jafari and H.~Johannesson, Phys.~Rev.~Lett. Phys.~Rev.~Lett. {\bf 118}, 015701 (2017).

 \bibitem{ubsb} U.~Bhattacharya, S.~Bandyopadhyay, and A.~Dutta, Phys.~Rev.~B {\bf 96}, 180303 (2017).

 \bibitem{ubad} U.~Bhattacharya and A.~Dutta, Phys.~Rev.~B {\bf 96}, 014302 (2017).

 \bibitem{utz} S.~Deffner and E.~Lutz, Phys.~Rev.~Lett. {\bf 105}, 170402 (2010).

 \bibitem{svaik}  S.~Vaikuntanathan and C.~Jarzynski, Europhys. Lett. {\bf 87}, 60005 (2009).

 \bibitem{lieb} E.~Lieb, T.~Schultz, and D.~Mattis, Ann.~Phys. (NY) {\bf 16}, 407 (1961).

 \bibitem{sachdev} S.~Sachdev, {\it Quantum Phase Transitions} (Cambridge University Press, Cambridge, UK, 1999).

 \bibitem{kos} P.~Kos and T.~Prosen, J.~Stat.~Mech. 123103 (2017).

 \bibitem{sharma} S.~Sharma and A.~Dutta, Phys.~Rev.~E {\bf 92}, 022108 (2015).

 \bibitem{sptja} S.~Paganelli and T.~J.~G.~Apollaro, Int.~J.~Mod.~Phys.~B {\bf 31}, 1750065 (2017).

 \bibitem{dorner1} R.~Dorner, S.~R.~Clark, L.~Heaney, R.~Fazio, J.~Goold,
                   and V.~Vedral, Phys.~Rev.~Lett. {\bf 110}, 230601 (2013).

 \bibitem{mazzola} L.~Mazzola, G.~De~Chiara, and M.~Paternostro, Phys.~Rev.~Lett. {\bf 110}, 230602 (2013).

 \bibitem{mazzola1} L.~Mazzola, G.~De~Chiara, and M.~Paternostro, Int.~J.~Quantum.~Inform. {\bf 12}, 1461007 (2014).

 \bibitem{tbam1} T.~B.~Batalhao, A.~M.~Souza, R.~S.~Sarthour, I.~S.~Oliveira
                 M.~Paternostro, E.~Lutz, and R.~M.~Serra, Phys.~Rev.~Lett. {\bf 115}, 190601 (2015).


 \end{thebibliography}
\end{document}